\documentstyle{article}
\begin{document}
\title
{
Bilinearization of Discrete Soliton Equations \\
and Singularity Confinement
}

\author
{
\renewcommand{\thefootnote}{\arabic{footnote}}
Kenichi Maruno$^{a,}$\footnotemark[1]
, Kenji Kajiwara$^{b}$
, Shinichiro Nakao$^{c}$
, Masayuki Oikawa$^{d}$\\
\small $^a$Department of Earth System Science and Technology, Kyushu University, 
Kasuga, Fukuoka 816\\
\small $^b$Department of Electrical Engineering, Doshisha University, Tanabe, 
Kyoto 610-03\\
\small $^c$Department of Applied Mathematics and Informatics, Ryukoku 
University, Seta, Otsu 520-21\\
\small $^d$Research Institute for Applied Mechanics, Kyushu University, Kasuga, 
Fukuoka 816\\}
\footnotetext[1]{E-mail: maruno@riam.kyushu-u.ac.jp.}
\date{Received \quad \quad October 1996}
\maketitle
\begin{abstract}
Bilinear forms for some nonlinear partial difference equations(discrete 
soliton equations) are derived based on the results of singularity 
confinement. Using the bilinear forms, the {\it N}-soliton and algebraic 
solutions of the discrete potential mKdV equation are constructed.
\end{abstract}

\section{Introduction}
 Hirota's method is one of the most powerful methods to study
soliton equations.  The crucial step in the method is to find
the proper dependent variable transformation (DVT) which 
reduces a given soliton equation to bilinear 
(or multilinear) forms, that is,
to express the dependent variable in terms of a $\tau$ function. 
Once we get the bilinear form, 
we are able to construct systematically particular solutions 
including the multisoliton solutions.

The bilinear approach is also useful for discrete soliton 
equations.  In order to obtain the integrable discrete versions 
of soliton equations,
Hirota discretized the bilinear forms of the soliton equations
so as to preserve solution structure\cite{rf:1,rf:2}.
Then finding a proper DVT, we are able to construct the nonlinear
difference equation.

But when we try to find the solutions for a given nonlinear
difference equation, it is difficult to find a
DVT which reduces a given discrete soliton equation to 
bilinear (or multilinear) forms, even if we know the 
results of corresponding continuous case.

Recently, Ramani, Grammaticos, and Satsuma proposed to use
the singularity confinement test to get proper DVT's for
nonlinear ordinary difference equations, and derived those for 
the discrete Painlev\'{e} equations\cite{rf:3}.   
Singularity confinement (SC) was introduced 
as the discrete analogue of the Painlev\'{e} property for continuous 
systems, that is, as a criterion for integrability of discrete
systems\cite{rf:4}.
In this article we apply  this method 
based on SC to some nonlinear partial difference equations 
to find the proper DVT's and the bilinear forms for them.  
In particular, for the discrete mKdV equation we construct 
the {\it N}-soliton and algebraic solutions. 
\section{Discrete KdV Equation}
We first consider the discrete KdV equation of the form\cite{rf:5}
\begin{equation}
  u_n^{m+1}-u_{n+1}^{m-1}=\frac{\epsilon }{\delta }
  \left(\frac{1}{u_n^m}-\frac{1}{u_{n+1}^m}\right).\label{eqn:a1}
\end{equation}
We have a singularity whenever the $u_n^m$ in the denominator 
happens to vanish. This has as a consequence that both $u_n^{m+1}$ 
and $u_{n-1}^{m+1}$ diverge,whereupon $u_{n-1}^{m+2}$ vanishes again 
and $u_{n-1}^{m+3},u_{n-2}^{m+3},u_{n-2}^{m+4}$ are finite. 
Thus the singularity is confined.  Following the SC criterion, 
therefore, the discrete KdV equation is integrable.
Singularity pattern of this case is 
\begin{equation}
\{u_n^m,u_n^{m+1},u_{n-1}^{m+1},u_{n-1}^{m+2}\}\to \{0,\infty,\infty,0\}.
\end{equation}
We use this information in order to express $u$ in terms of 
$\tau$ functions. 
Let us assume that the singularity pattern is caused by the fact that
a $\tau$ function $\tau_n^m$ becomes zero at $(m, n)$.  
Then, the singularity pattern suggests the transformation
\begin{equation}
   u_n^m=\frac{\tau_n^m\tau_{n+1}^{m-2}}
{\tau_n^{m-1}\tau_{n+1}^{m-1}}.\label{eqn:b1}
\end{equation}
Since $u_n^{m+1}$ and $u_{n-1}^{m+1}$ contain  $\tau_n^m$  in
the denominator, and  $u_n^m$ and  $u_{n-1}^{m+2}$ contain 
$\tau_n^m$  in the numerator, eq.(\ref{eqn:b1}) is certainly 
consistent with the singularity pattern.
Substituting eq.(\ref{eqn:b1}) into eq.(\ref{eqn:a1}) we obtain 
readily the trilinear form
\begin{equation}
\delta \tau_n^{m+1}\tau_{n+1}^{m-2}\tau_{n+2}^{m-2}
-\epsilon \tau_n^{m-1}\tau_{n+1}^m\tau_{n+2}^{m-2}
=\delta \tau_{n+2}^{m-3}\tau_n^m\tau_{n+1}^m
-\epsilon \tau_{n+2}^{m-1}\tau_n^m\tau_{n+1}^{m-2}.\label{eqn:a2}
\end{equation}
Muliplying eq.(\ref{eqn:a2}) by $\tau_{n+1}^{m-1}$ and 
manipulating the result, we obtain
\begin{equation}
F_n^m = F_{n+1}^{m-1}, \  
F_n^m \equiv \frac{\delta \tau_n^{m+1}\tau_{n+1}^{m-2}-
\epsilon \tau_n^{m-1}\tau_{n+1}^m}
{\tau_{n+1}^{m-1}\tau_n^m}.\label{eqn:b2}
\end{equation} 
This gives rise to $F_n^m = \alpha(m+n)$, here $\alpha(m+n)$ is
an arbitrary function of $m+n$.
Thus we obtain the bilinear form:
\begin{equation}
\delta \tau_n^{m+1}\tau_{n+1}^{m-2}-\epsilon \tau_n^{m-1}\tau_{n+1}^m
-\alpha(m+n)\tau_{n+1}^{m-1}\tau_n^m=0.
\end{equation}
In case of $\alpha(m+n)=1$, this bilinear form is the same
as that obtained by Hirota\cite{rf:5}.
From this we can obtain the solutions through formal perturbation
procedure.
\section{Discrete Toda Equation}
Secondly, we consider the discrete Toda equation of the form\cite{rf:5}
\begin{eqnarray}
  &&I_n^{m+1}-I_n^m=V_n^m-V_{n-1}^{m+1},\label{eqn:a3}\\
  &&I_n^{m+1}V_n^{m+1}=I_{n+1}^mV_n^m.\label{eqn:a4}
\end{eqnarray}
We have a singularity whenever the $V_n^m$ 
happens to vanish.
The singularity pattern is
\begin{eqnarray}
\{V_n^m,V_{n+1}^m,V_{n+1}^{m-1},V_{n+2}^{m-1}\}\to 
\{0,\infty,\infty,0\},\\
\{I_{n+1}^m,I_{n+1}^{m-1},I_{n+2}^m,I_{n+2}^{m-1}\}\to 
\{\infty,0,0,\infty\},
\end{eqnarray}
and the singularity is confined.
Let us assume again that the singularity pattern arises
from that a $\tau$ function $\tau_n^m$
vanishes at $(m, n)$.  Then,
the expressions for $V$ and $I$, dictated by the singularity pattern, are 
\begin{eqnarray}
  V_n^m&=&\frac{\tau_n^m\tau_{n-2}^{m+1}}
{\tau_{n-1}^m\tau_{n-1}^{m+1}},\\
  I_n^m&=&\frac{\tau_{n-1}^{m+1}\tau_{n-2}^m}
{\tau_{n-1}^m\tau_{n-2}^{m+1}}.
\end{eqnarray}
Substituting these into eqs.(\ref{eqn:a3}) and (\ref{eqn:a4}),
we obtain readily
\begin{equation}
  \tau_{n}^{m+1}\tau_{n}^{m-1}-\tau_{n+1}^{m-1}\tau_{n-1}^{m+1}
  =\alpha(m)[\tau_{n}^{m}]^2.
\end{equation}
Here, $\alpha(m)$ is an arbitrary function of $m$.
In case of $\alpha(m)=1$, the above bilinear form is the same
as that obtained in Ref.\cite{rf:5}.
\section{Discrete Hungry Lotka-Volterra Equation}
Next, we consider the discrete Hungry Lotka$-$Volterra equation 
of the form\cite{rf:6}
\begin{equation}
  \frac{u_n^{t+1}}{u_n^t}=
  \prod_{i=1}^M \frac{1+\delta u_{n-i}^t}
{1+\delta u_{n+i}^{t+1}}.\label{eqn:b4}
\end{equation}
Examining eq.(\ref{eqn:b4}) in the form solved with respect to 
$u_{n+M}^{t+1}$, we find the singularity pattern
\begin{equation}
\{u_{n-M-1}^{t-1},u_{n-1}^{t-1},u_{n-1}^t,u_n^{t-1},u_n^t,u_{n+M}^t 
\}\to \{0,\infty,-\frac{1}{\delta},-\frac{1}{\delta},\infty,0 \},
\end{equation}
and the singularity is confined.
However, when we assume the singularity pattern arises from 
vanishing of a $\tau$ function, 
we can't obtain the expression of $u$ in terms of the $\tau$ function
which satisfies the singularity pattern. So we divide the singularity pattern 
into two patterns:
\begin{eqnarray}
 &&\{u_{n-M-1}^{t-1},u_{n-1}^{t-1},u_n^t,u_{n+M}^t \}
\to \{0,\infty,\infty,0 \},\\
 &&\{u_{n-1}^{t-1},u_{n-1}^t,u_n^{t-1},u_n^t \}
\to \{ \infty,-\frac{1}{\delta},-\frac{1}{\delta},\infty \}.
\end{eqnarray}
These patterns suggest the transformations
\begin{eqnarray}
u_n^t
&=&\frac{\tau_{n+M+1}^{t+1}\tau_{n-M}^t}{\tau_{n+1}^{t+1}\tau_n^t}\\
&=&a\frac{\tau_{n+1}^{t}\tau_{n}^{t+1}}
{\tau_{n+1}^{t+1}\tau_n^t}-\frac{1}{\delta}, \ (a : \mbox{const}).
\end{eqnarray}
We thus obtain the bilinear equation of the form
\begin{equation}
  \delta \tau_{n+M+1}^{t+1}\tau_{n-M}^t
  -a \delta \tau_{n+1}^t\tau_n^{t+1}
  +\tau_{n+1}^{t+1}\tau_n^t=0,
\end{equation}
which is the same as that obtained in Ref.\cite{rf:6}.
\section{Discrete Potential mKdV Equation}
Let us next consider the discrete potential mKdV equation of 
the form\cite{rf:7}
\begin{equation}
  v_{n-1}^{m+2}=v_n^m\frac{\mu v_{n-1}^{m+1}-v_n^{m+1}}
{\mu v_n^{m+1}-v_{n-1}^{m+1}}.\label{eqn:c1}
\end{equation}
To apply SC, it is convenient to consider 
\begin{equation}
 I_n^{m+1}=I_{n+1}^{m-1}\frac{\mu -I_{n+1}^m}{\mu I_{n+1}^m-1}
  \frac{\mu I_n^m-1}{\mu -I_n^m},\label{eqn:b7} 
\end{equation}
where $I_n^m$ and $v_n^m$ are related as
$I_n^m=v_n^m/v_{n-1}^m$.
In this case we have two singularity patterns.
Either $I_n^m$ first takes the value $1/\mu$, in which case we have
\begin{equation}
\{I_n^m,I_n^{m+1},I_{n-1}^{m+1},I_{n-1}^{m+2}\}\to 
\{\frac{1}{\mu},0,\infty,\mu \},\label{eqn:a5}
\end{equation}
and the singularity is confined, or we start with $\mu$,
in which case we have
\begin{equation}
\{I_n^m,I_n^{m+1},I_{n-1}^{m+1},I_{n-1}^{m+2}\}\to 
\{\mu,\infty,0,\frac{1}{\mu}\},\label{eqn:a6}
\end{equation}
and the singularity is confined.
We assume that the singularity patterns (\ref{eqn:a5}) 
and (\ref{eqn:a6}) come from vanishing of  
$\tau$ functions $F_n^{m}$ and $G_n^{m}$ at $(n,m+1)$.
Then it is suggested that $I_n^m$ are in the forms
\begin{eqnarray}
I_n^m&=& \mu + \frac {F_{n+1}^{m-1}}{F_{n+1}^m}P\nonumber\\
&=& \frac{1}{\mu} + \frac {F_n^{m+1}}{F_{n+1}^m}Q\nonumber\\
&=& \frac{F_n^m}{F_{n+1}^m}R, \label{eqn:b5}
\end{eqnarray}
\begin{eqnarray}
I_n^m&=& \mu + \frac {G_n^{m+1}}{G_n^m}S\nonumber\\
&=& \frac{1}{\mu} + \frac {G_{n+1}^{m-1}}{G_n^m}T\nonumber\\
&=& \frac{G_{n+1}^m}{G_n^m}U, \label{eqn:b6}
\end{eqnarray}
where $P,Q,R$ must be expressed in terms of 
$G_n^m$'s and $S,T,U$ in terms of $F_n^m$'s.
Combining eqs.(\ref{eqn:b5}) and (\ref{eqn:b6}), 
we find the following simple 
expressions for $I_n^m$
\begin{eqnarray}
I_n^m&=& \mu + \alpha \frac {F_{n+1}^{m-1} G_n^{m+1}}{F_{n+1}^m
 G_n^m}\\
&=& \frac{1}{\mu} + \beta \frac {F_n^{m+1} G_{n+1}^{m-1}}{F_{n+1}^m 
G_n^m}\\
&=& \frac{F_n^m G_{n+1}^m}{F_{n+1}^mG_n^m},\label{eqn:b8}
\end{eqnarray}
where $\alpha,\beta$ are arbitrary constants.
The bilinear equations are obtained 
from the equivalence of the above three expressions of $I_n^m$.
\begin{eqnarray}
&&F_n^mG_{n+1}^m-\mu F_{n+1}^mG_n^m=\alpha 
F_{n+1}^{m-1}G_n^{m+1},\label{eqn:a7}\\
&&\mu F_n^mG_{n+1}^m-F_{n+1}^mG_n^m=\mu \beta 
F_n^{m+1}G_{n+1}^{m-1}.\label{eqn:a8}
\end{eqnarray}
The equation (\ref{eqn:b7}) is, then, satisfied automatically.
From the expression (\ref{eqn:b8}), we have
\begin{equation}
v_n^m = \frac{G_{n+1}^m}{F_{n+1}^m}.
\end{equation}
In order for this to satisfy (\ref{eqn:c1}), 
$\beta = -\alpha /\mu$  is required.
\section{Discrete mKdV Equation}
We finally consider the discrete mKdV equation proposed by
Tsujimoto and Hirota\cite{rf:10},
\begin{equation}
\frac{w_k^{l+1}(1+\delta w_{k+1}^{l+1})}{1+aw_k^{l+1}}
=\frac{w_k^{l}(1+\delta w_{k-1}^{l})}{1+aw_k^{l}}\label{dmKdV}
\ .
\end{equation}
Bilinerization of eq.(\ref{dmKdV}) is done in a similar way to
the previous section. We have two singularity patterns, either
$w_k^l$ takes the value $-\frac{1}{a}$ in which case we have
\begin{equation}
\{w_k^l,w_{k+1}^{l},w_{k+1}^{l+1},w_{k+2}^{l+1}\}\to 
\{-\frac{1}{a},-\frac{1}{\delta},\infty,0\},\label{dmKdV:p1}
\end{equation}
and the singularity is confined, 
or we start with $0$, in which case we have
\begin{equation}
\{w_k^l,w_{k+1}^{l},w_{k+1}^{l+1},w_{k+2}^{l+1}\}\to 
\{0,\infty,-\frac{1}{\delta},-\frac{1}{a}\},\label{dmKdV:p2}
\end{equation}
and the singularity is confined. 
From these patterns, we obtain the expression for $w_k^l$,
\begin{eqnarray}
w_k^l&=&
-\frac{1}{a}+\alpha^\prime\frac{\kappa_k^{l}\sigma_{k-2}^{l-1}}{\kappa_{k-1}^{l-
1}\sigma_{k-1}^l}\ 
,\label{dmKdV:dvt1}\\
&=&\beta^\prime\frac{\kappa_{k-2}^{l-1}\sigma_k^l}{\kappa_{k-1}^{l-1}\sigma_{k-1
}^l}\ 
,\label{dmKdV:dvt2}\\
&=&-\frac{1}{\delta}+\gamma^\prime\frac{\kappa_{k-1}^l\sigma_{k-1}^{l-1}}
{\kappa_{k-1}^{l-1}\sigma_{k-1}^l}\ ,\label{dmKdV:dvt3}
\end{eqnarray}
where $\alpha^\prime$, $\beta^\prime$ and $\gamma^\prime$ are
arbitrary constants.
Then we get the bilinear equations from the equivalence of these
expressions,
\begin{equation}
\delta\gamma^\prime\kappa_k^{l+1}\sigma_k^l-\kappa_k^l\sigma_k^{l+1}-\beta^\prime\delta
\kappa_{k-1}^{l}\sigma_{k+1}^{l+1}=0\ ,\label{bleq_dmKdV:2}
\end{equation}
\begin{equation}
\delta\gamma^\prime\kappa_k^{l+1}\sigma_k^l-\left(1-\frac{\delta}{a}\right)\kappa_k^l\sigma_k^{l+1}
-\alpha^\prime\delta\kappa_{k+1}^{l+1}\sigma_{k-1}^l=0\ .\label{bleq_dmKdV:3}
\end{equation}
Comparing these equations
with eqs.(\ref{eqn:a7}) and (\ref{eqn:a8}),
then we easily find that they are equivalent if we put
\begin{equation}
m=-k,\ n=l,\ F=\kappa, G=\sigma,\label{equiv:1}
\end{equation}
and
\begin{equation}
\gamma^\prime=\frac{1}{\delta \mu}
=\frac{1}{\delta}\left(1-\frac{\delta}{a}\right)^{\frac{1}{2}}\
,\quad\mu=\frac{1}{\delta\gamma^\prime},\quad
\alpha=-\frac{\delta \alpha^\prime}{1-\frac{\delta}{a}},\quad
\beta=-\frac{\beta^\prime}{\mu \gamma^\prime}=-\delta \beta^\prime .
\label{equiv:2}
\end{equation}
Finally, we obtain the explicit relationship between discrete 
potential mKdV equation and discrete mKdV equations. 
Noticing the relation eqs.(\ref{equiv:1}) and
(\ref{equiv:2}), we find from eq.(\ref{dmKdV:dvt3})that $w$ and $v$ are related 
as
\begin{equation}
w_m^n=\frac{1}{\delta}\left(\frac{1}{\mu}\frac{v_{n-2}^{m+1}}{v_{n-1}^{m+1}}-1\right)\ ,
\end{equation}
which can be also verified directly.
The point is that it requires ``art of discovery''to find
the above relationship just by looking at the equation, 
but now it becomes a ``routine work'' if we look at 
the bilinear equations which is obtained through SC.
\section{{\it N}-soliton Solution of the discrete potential mKdV Equation}
Let us construct the {\it N}$-$soliton solution of the discrete 
potential mKdV equation. Indeed, we can use the perturbative
technique to obtain the {\it N}-soliton solution, but here
we derive it from the reduction from the discrete KP equation.
The bilinear form of discrete KP equation is written as
\begin{eqnarray}
&&\quad a_1(a_2-a_3)\tau(n_1+1,n_2,n_3) \tau(n_1,n_2+1,n_3+1)\nonumber\\
&&+a_2(a_3-a_1)\tau(n_1,n_2+1,n_3) \tau(n_1+1,n_2,n_3+1)\nonumber\\
&&+a_3(a_1-a_2)\tau(n_1,n_2,n_3+1) \tau(n_1+1,n_2+1,n_3)=0,
\label{eqn:a9}
\end{eqnarray}
where $\tau$ depends on three discrete independent variables $n_1$, $n_2$ 
and $n_3$, and $a_1$,$a_2$ and $a_3$ are 
the difference intervals for $n_1$,$n_2$ and $n_3$, respectively
\cite{rf:8}.
The Casorati determinant solution to eq.(\ref{eqn:a9}) is given by\cite{rf:11}
\begin{equation}
 \tau (n_1,n_2,n_3)
=\left\vert \begin{array}{cccc}
  f_1(n_1,n_2,n_3;s) & f_1(n_1,n_2,n_3;s+1) & \cdots &
f_1(n_1,n_2,n_3;s+N-1) \\
  f_2(n_1,n_2,n_3;s) & f_2(n_1,n_2,n_3;s+1) & \cdots &
f_2(n_1,n_2,n_3;s+N-1) \\
\cdots & \cdots & \cdots & \cdots \\
  f_N(n_1,n_2,n_3;s) & f_N(n_1,n_2,n_3;s+1) & \cdots &
f_N(n_1,n_2,n_3;s+N-1) \\
  \end{array}\right\vert ,
\end{equation}
where $f_j$'s are arbitrary functions of $n_1$,$n_2$ and $n_3$ which satisfy
the dispersion relations
\begin{equation}
 \Delta _{n_k}f_j(n_1,n_2,n_3;s)=f_j(n_1,n_2,n_3;s+1) \quad
(k=1,2,3).\nonumber\label{dispersion}
\end{equation}
Here $\Delta_{n_k}$ are the backward difference operators:
\begin{equation}
  \Delta _{n_k}f(n_k)\equiv \frac{f(n_k)-f(n_k-1)}{a_k} \quad (k=1,2,3).
\end{equation}
For example,we can take $f_j$ as
\begin{eqnarray}
f_j(n_1,n_2,n_3;s)&=&p_j^s(1-p_j a_1)^{-n_1}(1-p_j a_2)^{-n_2}
(1-p_j a_3)^{-n_3}\nonumber\\
&\quad &+q_j^s(1-q_j a_1)^{-n_1}(1-q_j a_2)^{-n_2}
(1-q_j a_3)^{-n_3}, \label{eqn:b10}
\end{eqnarray}
where $p_j,q_j$ are arbitrary constants, then it will give the
{\it N}-soliton solution.

Now we assume that 
there exist a non-zero constant $\Phi$ such that 
for arbitrary $n_1,n_2,n_3$\cite{rf:9} 
\begin{equation}
\tau(n_1,n_2,n_3-2)=\Phi\tau(n_1,n_2,n_3).\label{eqn:c2}
\end{equation}
Then defining as 
$F(n_1,n_2)=\tau(n_1,n_2,n_3)$, $G(n_1,n_2)=\tau(n_1,n_2,n_3+1)$
we have  
\begin{eqnarray}
  &&\quad a_1(a_2-a_3)F(n_1+1,n_2)G(n_1,n_2+1)\nonumber\\
  &&+a_2(a_3-a_1)F(n_1,n_2+1)G(n_1+1,n_2)\nonumber\\
  &&+a_3(a_1-a_2)F(n_1+1,n_2+1)G(n_1,n_2)=0,\label{eqn:a15}
\end{eqnarray}
and by shifting $n_3 \to n_3+1$ in eq.(\ref{eqn:a9}) and 
using eq.(\ref{eqn:c2})   
\begin{eqnarray}
 &&\quad a_1(a_2-a_3)G(n_1+1,n_2)F(n_1,n_2+1)\nonumber\\
  &&+a_2(a_3-a_1)G(n_1,n_2+1)F(n_1+1,n_2)\nonumber\\
  &&+a_3(a_1-a_2)G(n_1+1,n_2+1)F(n_1,n_2)=0.\label{eqn:a16}
\end{eqnarray}
After shifting $n_1\to n_1-1$,eqs.(\ref{eqn:a15}) and (\ref{eqn:a16}) 
are transformed into the following equation  
\begin{eqnarray}
&&a_1(a_2-a_3)F_n^mG_{n+1}^m+a_2(a_3-a_1)F_{n+1}^mG_n^m
=a_3(a_2-a_1)F_{n+1}^{m-1}G_n^{m+1},\label{eqn:d1}\\
&&a_2(a_3-a_1)F_n^mG_{n+1}^m+a_1(a_2-a_3)F_{n+1}^mG_n^m
=a_3(a_2-a_1)F_n^{m+1}G_{n+1}^{m-1},\label{eqn:d2}
\end{eqnarray}
through the variable transformation 
\begin{eqnarray}
  m&=&-n_1-n_2,\nonumber\\
  n&=&n_2,\label{eqn:d3}
\end{eqnarray}
where $F(n_1,n_2)$, $G(n_1,n_2)$ are defined by
\begin{equation}
F(n_1,n_2) \equiv F_n^m = F_{n_2}^{-n_1-n_2},
G(n_1,n_2) \equiv G_n^m = G_{n_2}^{-n_1-n_2}.
\label{eqn:d4}   
\end{equation}
Moreover if we must choose as 
\begin{equation}
\mu=-\frac{a_2(a_3-a_1)}{a_1(a_2-a_3)} ,
\alpha=\frac{a_3(a_2-a_1)}{a_1(a_2-a_3)} ,
\beta = \frac{a_3(a_2-a_1)}{a_2(a_3-a_1)},  
\end{equation}  
eqs.(\ref{eqn:a7}) and (\ref{eqn:a8}) reduce to eqs.(\ref{eqn:d1}) and (\ref{eqn:d2}).

Now we impose some constraint
on the parameters of the solution so that the condition
(\ref{eqn:c2}) is satisfied. We observe that
\begin{eqnarray}
&&f_j(n_1,n_2,n_3-2;s)\nonumber\\
  &&=p_j^s(1-p_j a_1)^{-n_1}(1-p_j a_2)^{-n_2}(1-p_j a_3)^{-n_3+2}
+q_j^s(1-q_j a_1)^{-n_1}(1-q_j a_2)^{-n_2}(1-q_j a_3)^{-n_3+2}\nonumber\\
  &&=p_j^s(1-p_j a_1)^{-n_1}(1-p_j a_2)^{-n_2}(1-p_ja_3)^{-n_3+2}\nonumber\\
&&
\times\left[ 1+ 
\left(\frac{q_j}{p_j}\right)^s\left(\frac{1-q_ja_1}{1-p_ja_1}\right)^{-n_1}
\left(\frac{1-q_ja_2}{1-p_ja_2}\right)^{-n_2}
\left(\frac{1-q_ja_3}{1-p_ja_3}\right)^{-n_3}
\left(\frac{1-q_ja_3}{1-p_ja_3}\right)^2\right]\ .
\end{eqnarray}
Thus, if we put
\begin{equation}
1-a_3 p_j=-(1-a_3 q_j)\ ,
\end{equation}  
or
\begin{equation}
q_j=\frac{2}{a_3}-p_j\ ,
\end{equation}
then we have
\begin{equation}
f_j(n_1,n_2,n_3-2;s)=(1-p_ja_3)^2 f_j(n_1,n_2,n_3;s)\ .
\end{equation} 
Using this relation we readily find the following relation:
\begin{eqnarray}
\tau(n_1,n_2,n_3-2)
=\prod_{k=1}^N(1-p_k a_3)^2~\tau(n_1,n_2,n_3)\ ,
\end{eqnarray}
that is, the condition (\ref{eqn:c2}) is satisfied for this solution.

Thus through the variable transformation (\ref{eqn:d3}) and (\ref{eqn:d4}),
the {\it N}-soliton solution of discrete potential mKdV equation 
(\ref{eqn:c1}) 
is given by 
\begin{eqnarray}
  &&v_n^m=\frac{G_{n+1}^m}{F_{n+1}^m},\\
  &&F_n^m
  =\left\vert \begin{array}{cccc}
  \phi _1(m,n;s) & \phi_1(m,n;s+1) & \cdots 
& \phi_1(m,n;s+N-1) \\
  \phi _2(m,n;s) & \phi_2(m,n;s+1) & \cdots 
& \phi_2(m,n;s+N-1) \\
  \cdots & \cdots & \cdots & \cdots \\
  \phi _N(m,n;s) & \phi_N(m,n;s+1) & \cdots 
& \phi_N(m,n;s+N-1) \\
  \end{array}\right\vert,\\
\nonumber  \\
  &&G_n^m
  =\left\vert \begin{array}{cccc}
  \psi _1(m,n;s) & \psi_1(m,n;s+1) & \cdots 
& \psi_1(m,n;s+N-1) \\
  \psi _2(m,n;s) & \psi_2(m,n;s+1) & \cdots 
& \psi_2(m,n;s+N-1) \\
  \cdots & \cdots & \cdots & \cdots \\
  \psi _N(m,n;s) & \psi_N(m,n;s+1) & \cdots 
& \psi_N(m,n;s+N-1) \\
  \end{array}\right\vert,
\end{eqnarray}
\begin{eqnarray}
  \phi _j(m,n;s)&=&
  p_j^s(1-p_j a_1)^{m+n}(1-p_j a_2)^{-n}
 +q_j^s(1-q_j a_1)^{m+n}(1-q_j a_2)^{-n},\\
  \psi_j(n_1,n_2;s)&=&
 p_j^s(1-p_j a_1)^{m+n}(1-p_j a_2)^{-n}
 -q_j^s(1-q_j a_1)^{m+n}(1-q_j a_2)^{-n},\\
q_j&=&\frac{2}{a_3}-p_j.
\end{eqnarray}
For example, the 1-soliton solution is given by 
\begin{equation}
v_n^m=
\frac{p_1^s(1-p_1a_1)^{m+n+1}(1-p_1a_2)^{-n-1}
-q_1^s(1-q_1a_1)^{m+n+1}(1-q_1a_2)^{-n-1}}
{p_1^s(1-p_1a_1)^{m+n+1}(1-p_1a_2)^{-n-1}
+q_1^s(1-q_1a_1)^{m+n+1}(1-q_1a_2)^{-n-1}}.
\end{equation}
\section{Algebraic Solution of the Potential mKdV Equation}
Let us construct the algebraic solution of 
the discrete potential mKdV equation through
the reduction of the discrete KP equation.

First notice that 
\begin{eqnarray}
f_j(n_1,n_2,n_3;s)&=&\frac{\partial}{\partial p_j}
p_j^s(1-p_j a_1)^{-n_1}(1-p_j a_2)^{-n_2}(1-p_j a_3)^{-n_3}\
,\nonumber\\
&=& \left(\frac{s}{p_j}+\sum_{k=1}^3\frac{n_ka_k}{1-p_ja_k}\right)
p_j^s(1-p_j a_1)^{-n_1}(1-p_j a_2)^{-n_2}(1-p_j a_3)^{-n_3}\ ,
\end{eqnarray}
satisfies 
the dispersion relation (\ref{dispersion}), and hence
this also give the solution of the discrete KP equation
(\ref{eqn:a9}). Moreover, if we notice that
the discrete KP equation (\ref{eqn:a9}) is invariant
under multiplication of $c_1q_1^{n_1}c_2q_2^{n_2}c_3q_3^{n_3}$
on $\tau$ function, where $c_j$ and $q_j$, $j=1,2,3$ are arbitrary
constants, we find that it gives the algebraic solution of the
discrete KP equation. More generally, $f_j$ can be chosen as
\begin{equation}
f_j(n_1,n_2,n_3)=\frac{\partial^{k_j}}{\partial p_j^{k_j}}
p_j^s(1-p_j a_1)^{-n_1}(1-p_j a_2)^{-n_2}(1-p_j a_3)^{-n_3}\ ,
\end{equation}
where $k_j$ is an arbitrary integer. In particular, we define
the polynomials $P_k$ by
\begin{equation}
\displaystyle \sum_{k=0}^{\infty} P_k(n_1,n_2,n_3)\lambda^k
=(1-\lambda a_1)^{-n_1}(1-\lambda a_2)^{-n_2}(1-\lambda a_3)^{-n_3}
\quad (P_k = 0, \quad {\rm for}\ k<0).
\end{equation}
The simplest examples of $P_k$ are given by
\begin{eqnarray*}
P_0&=&1,\\
P_1&=&a_1n_1+a_2n_2+a_3n_3,\\
P_2&=&\frac{1}{2}(a_1n_1+a_2n_2+a_3n_3)^2
+\frac{1}{2}(a_1^2n_1+a_2^2n_2+a_3^2n_3),\\
&&\cdots
\end{eqnarray*}
Then, the $\tau$ function
\begin{equation}
  \tau _N(n_1,n_2,n_3)
  =\left\vert \begin{array}{cccc}
  P_{i_1}(n_1,n_2,n_3) & P_{i_1+1}(n_1,n_2,n_3) & \cdots 
& P_{i_1+N-1}(n_1,n_2,n_3) \\
  P_{i_2-1}(n_1,n_2,n_3) & P_{i_2}(n_1,n_2,n_3) & \cdots 
& P_{i_2+N-2}(n_1,n_2,n_3) \\
  \cdots & \cdots & \cdots & \cdots \\
  P_{i_N-N+1}(n_1,n_2,n_3) & P_{i_N-N+2}(n_1,n_2,n_3) & \cdots 
& P_{i_N}(n_1,n_2,n_3) \\
  \end{array}\right\vert,\label{eqn:e3}
\end{equation}
where $i_1\geq i_2\geq\cdots i_N>0$ are integers,
is nothing but the Schur function solution of eq.
(\ref{eqn:a9}) labeled by the Young diagram $(i_1,i_2,\cdots,i_N)$.

In order to satisfy the condition (\ref{eqn:c2}), let us
consider the $\tau$ function,
\begin{equation}
  \tau _N(n_1,n_2,n_3)
  =\left\vert \begin{array}{cccc}
  P_{N}(n_1,n_2,n_3) & P_{N+1}(n_1,n_2,n_3) & \cdots 
& P_{2N-1}(n_1,n_2,n_3) \\
  P_{N-2}(n_1,n_2,n_3) & P_{N-1}(n_1,n_2,n_3) & \cdots 
& P_{2N-3}(n_1,n_2,n_3) \\
  \cdots & \cdots & \cdots & \cdots \\
  P_{-N+2}(n_1,n_2,n_3) & P_{-N+3}(n_1,n_2,n_3) & \cdots 
& P_{1}(n_1,n_2,n_3) \\
  \end{array}\right\vert,\label{eqn:e4}
\end{equation}
and impose a condition,
\begin{equation}
P_k(n_1,n_2,n_3-2)=P_k(n_1,n_2,n_3)-c^2 P_{k-2}(n_1,n_2,n_3),
\label{eqn:e1}
\end{equation}
where $c$ is some constant.
Then it is easily verified that the $\tau$ function satisfy
the condition (\ref{eqn:c2}).

For simplicity, we introduce a function,
\begin{equation}
\theta (n_1,n_2,n_3)
=(1-\lambda a_1)^{-n_1}(1-\lambda a_2)^{-n_2}(1-\lambda a_3)^{-n_3},
\end{equation}
which is a generating function of the polynomials $P_k$.
Then eq.(\ref{eqn:e1}) implies that
\begin{equation}
\theta (n_1,n_2,n_3-2)=\theta (n_1,n_2,n_3)-c^2\lambda^2
\theta (n_1,n_2,n_3)\ ,
\end{equation}
from which we have
\begin{equation}
1-\lambda a_3=\pm (1-c^2\lambda^2)^{\frac{1}{2}} \ .
\end{equation} 
Defining as 
$F(n_1,n_2)=\tau_N(n_1,n_2,n_3),G(n_1,n_2)=\tau_N(n_1,n_2,n_3+1)$
we have bilinear forms (\ref{eqn:a15}),(\ref{eqn:a16}) 
of discrete potential mKdV equation.
Thus through the variable transformation (\ref{eqn:d3}),
(\ref{eqn:d4})
the algebraic solution of discrete potential mKdV equation 
(\ref{eqn:c1}) is given by
\begin{eqnarray}
  &&v_n^m=\frac{G_{n+1}^m}{F_{n+1}^m},\\
  &&F_n^m
=\left\vert \begin{array}{cccc}
  P_N(m,n;n_3) & P_{N+1}(m,n;n_3) & \cdots 
& P_{2N-1}(m,n;n_3) \\
  P_{N-2}(m,n;n_3) & P_{N-1}(m,n;n_3) & \cdots 
& P_{2N-3}(m,n;n_3) \\
  \cdots & \cdots & \cdots & \cdots \\
  P_{-N+2}(m,n;n_3) & P_{-N+1}(m,n;n_3) & \cdots 
& P_1(m,n;n_3) \\
  \end{array}\right\vert,\\
\nonumber  \\
  &&G_n^m
=\left\vert \begin{array}{cccc}
  P_N(m,n;n_3+1) & P_{N+1}(m,n;n_3+1) & \cdots 
& P_{2N-1}(m,n;n_3+1) \\
  P_{N-2}(m,n;n_3+1) & P_{N-1}(m,n;n_3+1) & \cdots 
& P_{2N-3}(m,n;n_3+1) \\
  \cdots & \cdots & \cdots & \cdots \\
  P_{-N+2}(m,n;n_3+1) & P_{-N+1}(m,n;n_3+1) & \cdots 
& P_1(m,n;n_3+1) \\
  \end{array}\right\vert,
\end{eqnarray}
\begin{equation}
\displaystyle \sum_{k=0}^{\infty} P_k(m,n;n_3)\lambda^k
=(1-\lambda a_1)^{m+n}(1-\lambda a_2)^{-n}
(1-\lambda ^2 c ^2)^{-\frac{n_3}{2}}.
\end{equation}
Concretely,
\begin{eqnarray*}
P_0(m,n;n_3)&=&1,\\
P_1(m,n;n_3)&=&a_1m+(a_1-a_2)n,\\
P_2(m,n;n_3)&=&\frac{1}{2}a_1^2 m(m-1)+a_1(a_1-a_2)mn
+\frac{1}{2}(a_1-a_2)^2 n^2
-\frac{1}{2}(a_1+a_2)(a_1-a_2)n
-\frac{1}{2}c^2 n_3,\\
&\quad&\cdots
\end{eqnarray*}
where $n_3$ is an arbitrary constant.
\section{Conclusion}
In this article, we have derived dependent variable transformations 
for discrete soliton equations to reduce them to bilinear forms by 
using singularity confinement.
We have also  constructed the {\it N}-soliton solution of discrete mKdV 
equation.
We expect that this method based on SC is effective for study on 
other discrete nonlinear equations. 

One of the authors (K.K) was supported by 
the Grant-in-aid for Encouragement of Young Scientist, 
The Ministry of Education, Science, Sports and Culture, No.08750090.

\end{document}